# Self-Healing Effects in OAM Beams Observed on a 28 GHz Experimental Link

**Marek Klemes, Lan Hu, Greg Bowles, Mohammad Akbari, Soulideth Thirakoune, Michael Schwartzman, Kevin Zhang, Tan Huy Ho, David Wessel and Wen Tong**

Canada Research Center, Huawei Technologies Canada Co., Ltd., Kanata, Ontario, CANADA.

Corresponding author: (e-mail: marek.klemes@huawei.com).

This work was supported in part by the Canada Research Center, Huawei Technologies Canada Co., Ltd., Kanata, Ontario, CANADA.

**ABSTRACT** In this paper we document for the first time some of the effects of self-healing, a property of orbital-angular-momentum (OAM) or vortex beams, as observed on a millimeter-wave experimental communications link in an outdoors line-of-sight (LOS) scenario. The OAM beams have a helical phase and polarization structure, and have conical amplitude shape in the far field. The Poynting vectors of the OAM beams also possess helical structures, orthogonal to the corresponding helical phase-fronts. Due to such non-planar structure in the direction orthogonal to the beam axis, OAM beams are a subset of "structured light" beams. Such structured beams are known to possess self-healing properties when partially obstructed along their propagation axis, especially in their near fields, resulting in partial reconstruction of their structures at larger distances along their beam axis. Various theoretical rationales have been proposed to explain, model and experimentally verify the self-healing physical effects in structured optical beams, using various types of obstructions and experimental techniques. Based on these models, we hypothesize that any self-healing observed will be greater as the OAM order increases. Here we observe the self-healing effects for the first time in structured OAM radio beams, in terms of communication signals and channel parameters rather than beam structures. We capture the effects of partial near-field obstructions of OAM beams of different orders on the communications signals, and provide a physical rationale to substantiate that the self-healing effect was observed to increase with the order of OAM, agreeing with our hypothesis.

**INDEX TERMS** Self-healing, orbital angular momentum (OAM), millimeter waves, vortex beam, helical phase, helical Poynting vector, partial beam obstruction, structured light, conical beam, Butler matrix, DFT matrix, phase gradient, phase velocities, transverse wave vector, error-vector magnitude (EVM)

## I. INTRODUCTION

As the demand for wireless services continues to grow, successive generations of radio transmission technology have evolved to accommodate increasing numbers of users and their volumes of data. Refinements in the exploitation of the familiar time and frequency dimensions of wireless signals have begun to approach the limits of their capacity to deliver the required volumes of data at the rates and qualities of the growing demand, so by the 5-th generation (5G) the spatial dimension was opened to create even more user channels, in the form of multiple-antenna systems collectively known as multi-input multi-output (MIMO), which relied on the scattering environment to form diverse spatial paths from transmitter to one or more receivers. This included the two possible dimensions of polarization which have been in use for some time as means of increasing the channel capacity of fixed microwave backhaul links as their frequency allocations began to be exhausted.

Work is now well under way on the 6-th generation of wireless technology (6G) as demand for connectivity grows from connected personal users to connected sensors, automobiles and the wireless internet of things (IOT), and along with it, the search for more signal dimensions continues. It has been found that the phase structure of wireless signal waves can be open to more variety beyond that of plane waves, especially at the higher frequencies where antennas occupy less space and their beams are more controllable. Specifically in the field of optics, it has been found some 32 years ago that light can carry orbital angular momentum (OAM), in addition to the well-known spin angular momentum (SAM, also known as polarization) [1]. The OAM manifests as a linear progression of electrical phase of the beam around its axis, completing an integer number, *l,* of 360-degree cycles of phase in one physical 360-degree orbit around the axis of propagation. Its far-field amplitude pattern has a conical shape, as shown in Figure 1. Its phase has a helical pattern, the equi-phase contours forming spirals around the beam axis. In the physics and optics community, OAM was typically effected on laser light by means of a spiral dielectric phase plate whose thickness varied as a linear function of azimuth angle such that the slope of the phase delay was an integer number of degrees per degree of azimuth angle. (It turns out that some quarter-century before that, a similar effect was in use in radio-frequency (RF) circular antenna arrays, but radiating mostly in the azimuth plane, and was known as phase-modes [2].) As there is no theoretical limit to the number of phase cycles around the beam axis, OAM became another signal dimension open to exploitation in 6G wireless technology development. Other structured beams and their properties, already studied in the optics community, have also come into consideration for 6G wireless applications.

One appealing property of structured beams is that of self-reconstruction, or "self-healing" of its structure when it is partially obstructed close to its source aperture [3]. This sets them apart from other MIMO antenna radiations which rely on a rich scattering environment to effect their spatial diversity, but also incur obstruction losses. Having been characterizing a mm-wave OAM multiplexing wireless link as the subject of previous publications [4], it was of interest to us to observe any evidence of self-healing of the OAM beams, from the point of view of the communications link parameters rather that the physical structures of the beams as has mostly been done in the optical community.

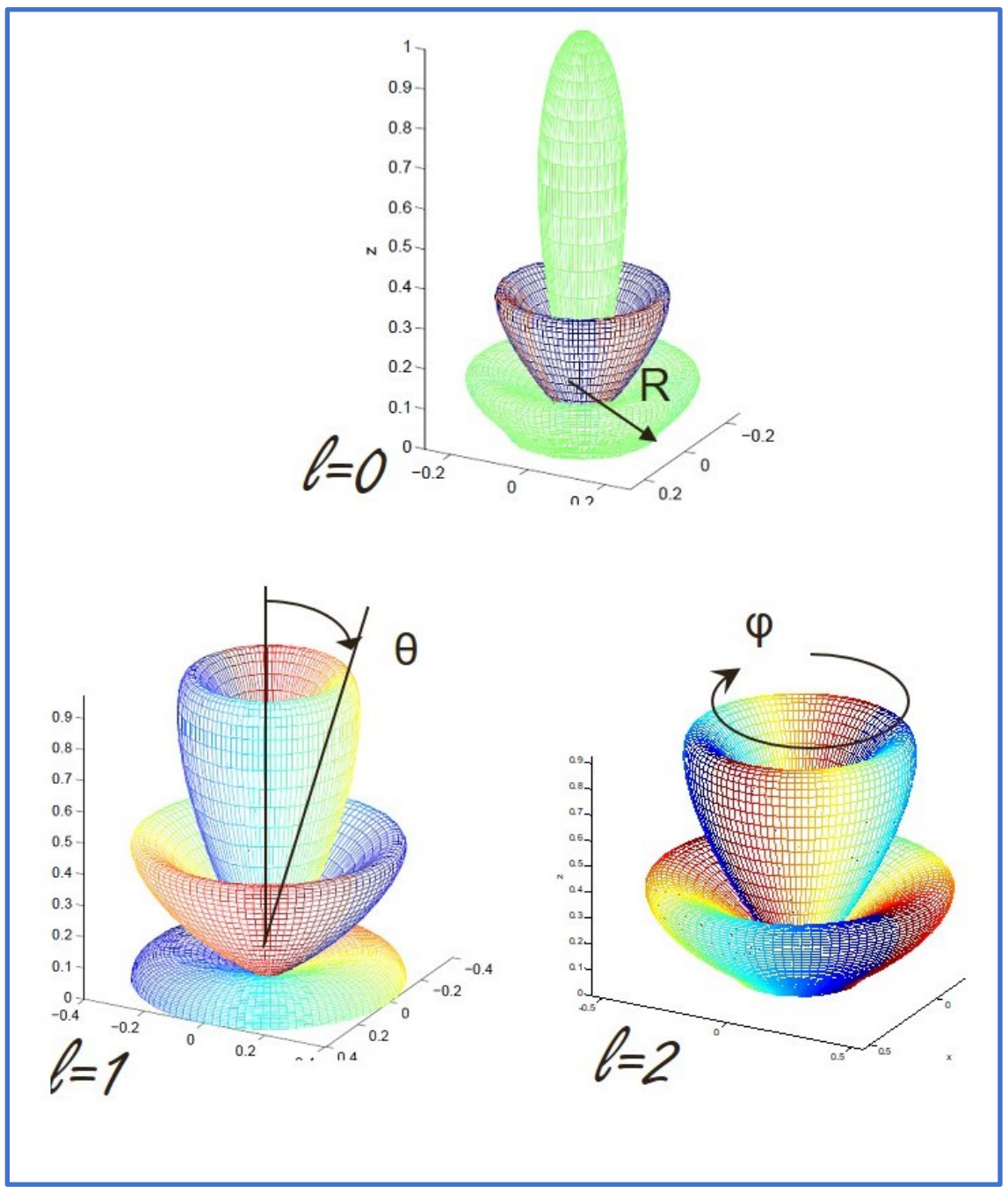

**Figure 1. Examples of OAM beams of order *l* = 0, +1, +2 generated by a uniform circular array. Colour denotes electrical phase, amplitude is normalized to peak value and popagation is along z axis (vertical).**

The paper is organized as follows: Section II summarizes the experimental mm—wave link, especially the multi-OAM transmitting antenna and its design rationale. Section III describes the self-healing experiments and summarizes the measured link performance data. Section IV analyzes the measured data and presents a physical rationale to account for the observed results, while section V draws the conclusions and outlines areas for future investigations.

## II. EXPERIMENTAL OAM OUTDOOR LINK AND OAM ANTENNA DESIGN

The OAM experimental test bench was designed as a one-way line-of-sight (LOS) link with the multi-mode OAM antenna transmitting an independent data stream on each of up to 5 mutually-orthogonal OAM beams to one multi-receiver user equipment (UE) fitted with a linear array of up to 4 antennas and corresponding receivers. More details of the original configuration are contained in reference [4]; some details were re-configured for purposes of the present experiments, as detailed in Table 1. Layout of the link and the equipment at each end is depicted in Figure 2. To keep sizes manageable, the RF was selected as 28 GHz, for a typical link length of 50m, with a far field ranging from 30m to 200m. To minimize perturbations of the OAM beams, the polarization was chosen as right-hand circular polarization (RHCP) at both ends, which would help to reject ground reflections, as all the equipment was mounted on movable carts at less than 2m from the ground. Accordingly, it was desired to keep the radius, *R*, of the conical OAM beams at the receiver to ~2m, so that the receiving (RX) antenna array could capture a significant portion of the beams, which in turn determined the size of the transmitting (TX) OAM antenna.

**Table 1. Experimental OAM link and TX antenna parameters**

| Parameter | Symbol | Relation | Value | Units |
|---|---|---|---|---|
| **Desired link specification.** | | | | |
| **Signal bandwidth** | *B* | (LTE channel) | 20 | MHz |
| **Number of OAM modes** | *K* | Co-channel data streams | 5 | 1 |
| **Link distance** | *L* | TX to RX | 50 | m |
| **RX antenna separation** | *d* | Max. UE antenna spacing | 0.20 | m |
| **Digital IF of RX** | *F* | ADC bandwidth | 983.04 | MHz |
| **Dependent link parameters** | | | | |
| **OAM beam radius** | *R* | $R< dF/(2B)$ $R=0.87m$ | 2.4576 | m |
| **RF wavelength** | $\lambda$ | $28.0x10^9$ $/3.0x10^8$ | 0.011 | m |
| **TX antenna radius (max).** | *r* | $r\sim(K-1)\lambda L/(4\pi R)$ | 0.201 | m |
| **TX far field** | $L_{far}$ | $2(2r)^2/\lambda$ | 32.35 | m |
| **No. of TX array elements (max)** | *N* | $N\sim 2\pi r/(\lambda/2)$ | 238 | 1 |

Additionally, all of the OAM beams needed to have the same cone radii so that they would overlap at the receiving end, because each beam carried a separate data stream on the same common RF carrier frequency, which needed to be de-multiplexed in the receiver. Therefore the TX antenna was composed of 3 concentric circular-array antennas, with the larger ones transmitting the correspondingly higher orders of OAM beams.

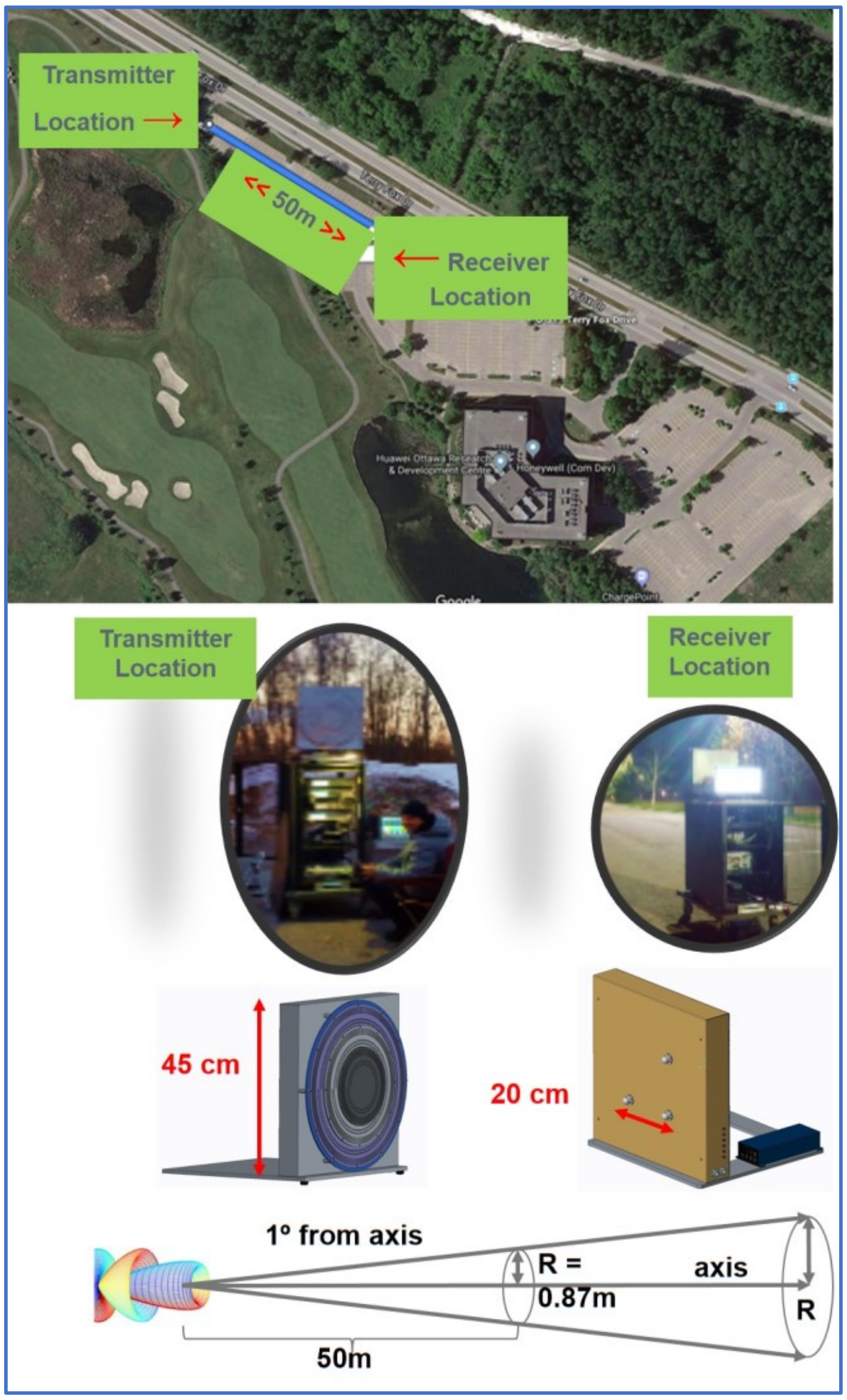

**Figure 2. Layout of experimental outdoor OAM link and 28 GHz transmitting and receiving equipment.**

Each array is a type of radial line slot array (RLSA), fed with a small circular array of probes around the center of the radial waveguide. Figure 3 depicts TX the multi-mode OAM antenna design. Its far-field radiation pattern of the $\ell$-th OAM mode can be analytically described by the equation

$$F_\ell(\theta,\varphi) = (-j)^\ell e^{j\ell\varphi} K J_\ell\left(\frac{2\pi r \sin\theta}{\lambda}\right) \tag{1}$$

where $r$ is the radius of the transmitting circular (actually spiral) array radiating the $\ell$-th OAM mode, $\lambda$ is the radiating wavelength, $J_\ell$ is the $\ell$-th order Bessel function of the first kind, $K$ is a constant of propagation, $\theta$ is the elevation or tilt angle from the beam ($z$) axis while $\varphi$ is the azimuth angle around the beam axis as well as the electrical phase of the beam at that azimuth angle in the far field.

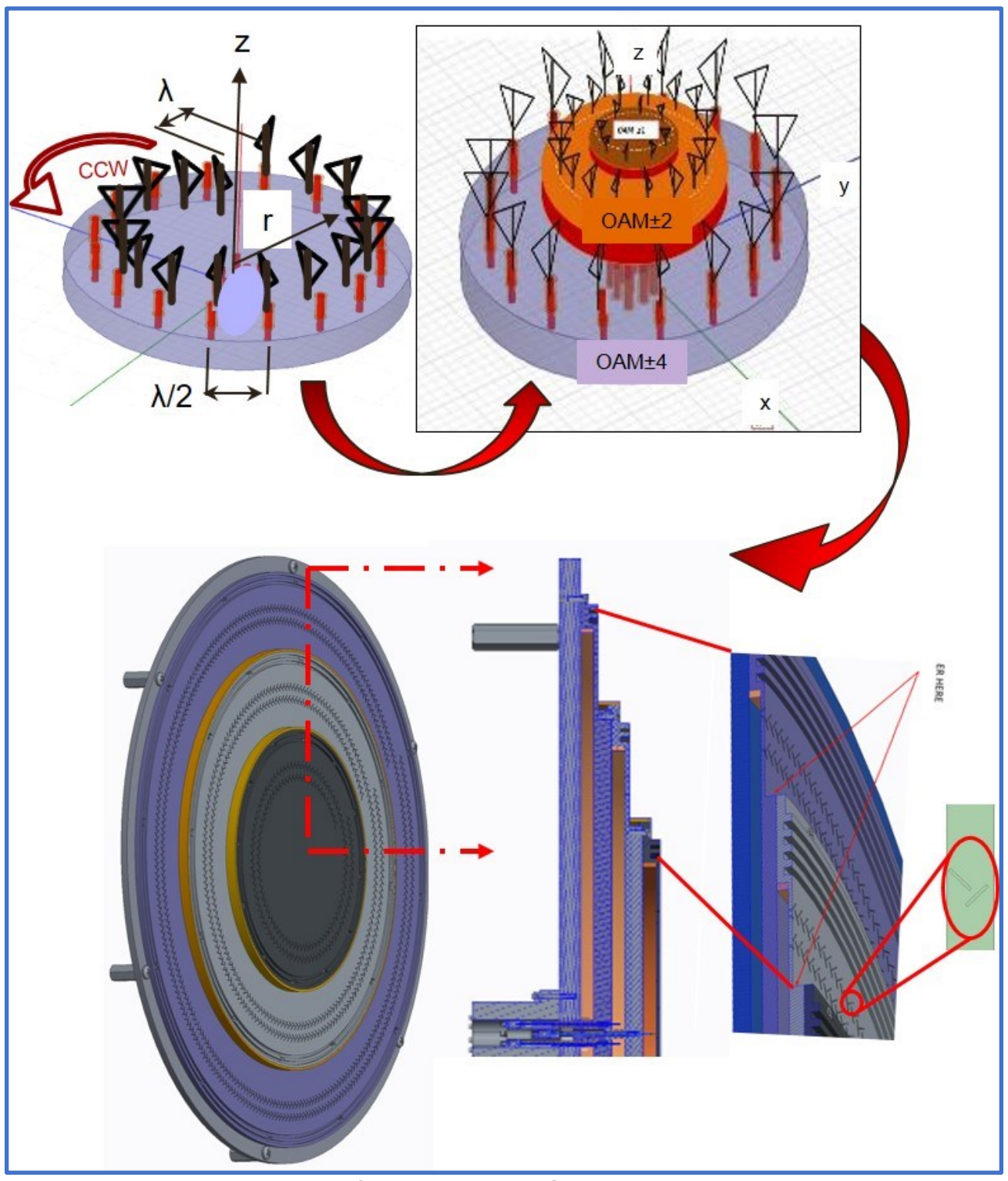

**Figure 3. Design of multi-mode OAM antenna, showing radially-oriented RHCP array elements, arranged in CCW outward spiral to compensate rotation phase of RHCP, stacked in discs of increasing radii for increasing OAM orders, with cross-section detail and detail of RHCP crossed-slots array elements.**

The receiving antenna array was positioned at a distance of $L$=50m from the transmitting OAM antenna, which was tilted upward so as to place the receiver at the bottom arc of the 1° cone of the OAM+2 beam transmitted from the middle RLSA and also the 1° cone of OAM+4 transmitted from the largest RLSA.

The radial profiles of these two OAM beams follow equation (1), each RLSA radius $r$ being designed so that the amplitude patterns of these two OAM beams coincide in the far field with $r_{OAM2}$ = 0.149m and $r_{OAM4}$ = 0.218m, as illustrated in Figure 4. Note that the diameters of the OAM+2 (green curve and outlined inset), and the scaled OAM+4 (red curve and outlined inset) are now both equal, corresponding to a cone angle of about 1.43 degrees from the beam axis. The receiver array was actually placed somewhat inside these conical beams, at $\theta$ ~1.0 ° because the phase gradients with respect to azimuth angle $\varphi$ were less rippled there, according to electromagnetic simulations of the OAM antenna. This scaling of the OAM amplitude profiles then leaves their ***azimuth (or circumferential) phase profiles as the main*** ***structural difference between the OAM+2 and OAM+4 beams, to which any self-healing may be attributed.***

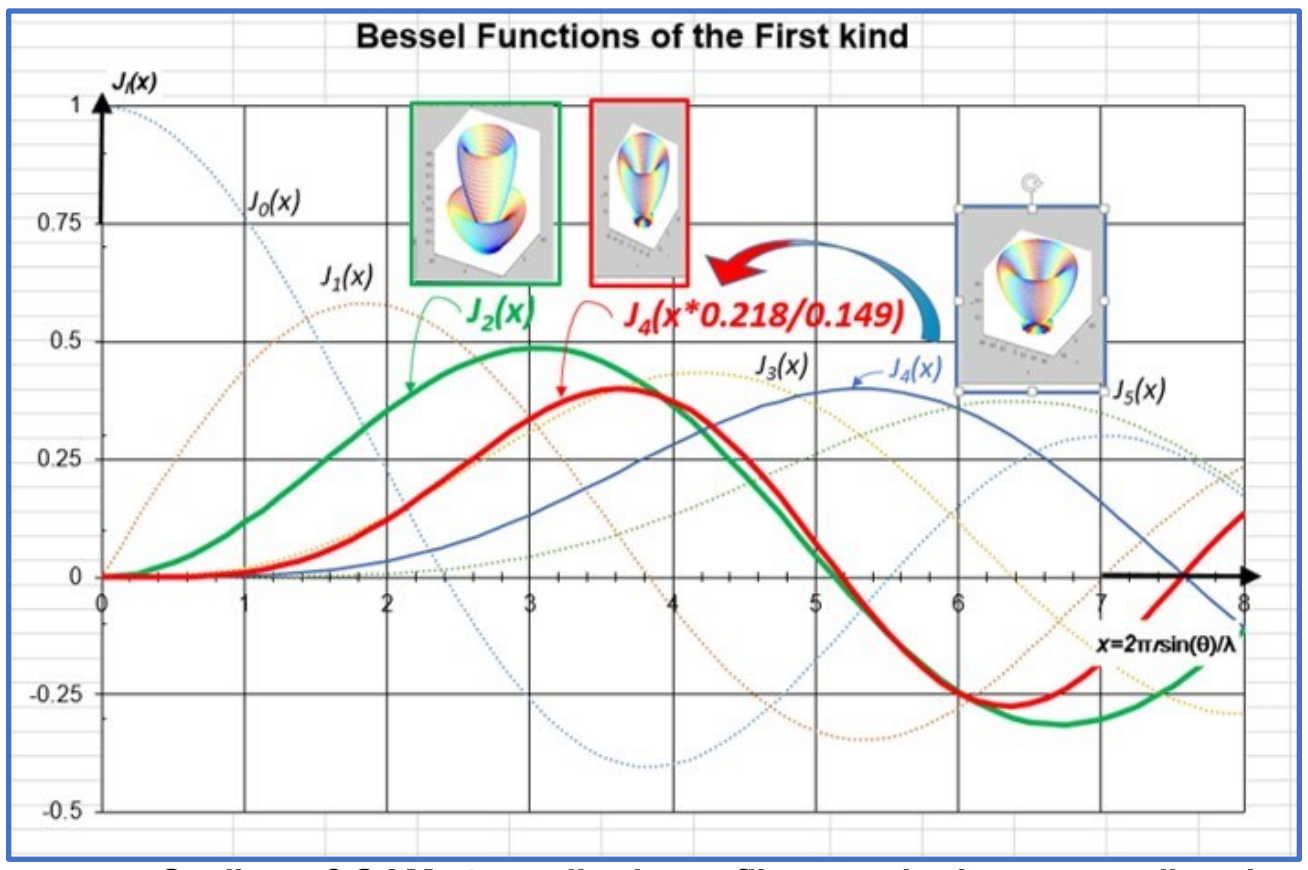

Figure 4. **Scaling of OAM+4 amplitude profile to make its cone radius the same as cone radius of OAM+2 in the far field. OAM+2 is sourced from aperture of diameter 0.149m (green) so OAM+4 (blue) must be sourced from aperture of diameter 0.218m (red) to overlap with OAM+2 beam at the receiver.**

This is very important, because we want to avoid causing similar effects as signs of self-healing, by other effects not related to their phase structuring, such as the differences in angular size of the central nulls and in the peak directivities of the OAM beams. Transmitted signal levels were adjusted to be equal in the OAM+2 and OAM+4 beams at the receiver.

## III. EFFECTS OF PARTIAL BEAM OBSTRUCTION AND SELF-HEALING IN OAM+2 AND OAM+4 BEAMS

In order to observe the effects of partial beam obstruction and discern that self-healing is taking place in the OAM beams, only relative measurements could be made, as the experimental link did not possess equipment capable of characterizing the entire beam structures. Specifically, the types of relative measurements of interest pertained to the communications parameters of the LOS link, much like in a MIMO link, such as signal-to-noise ratio (SNR), EVM, and phase gradient among the 4 antennas in the receiving array.

According to findings in recent literature, the self-healing effect in structured EM beams is typically evident in the presence of partial obstructions in their near fields [5], and significantly dependent on the wavelengths, especially in the region of mmm-waves. It is also not complete healing, depending on the relative size and position of the partial obstruction, which makes it difficult to quantify analytically [5], [6]. It is generally evident that structured beams which possess the largest proportions of wave vectors which have different directions from that in the direction of beam propagation (along its optical axis, $z$), also exhibit the greatest degrees of self-healing and structural robustness [5-7].

Our hypothesis in these experiments is that OAM+2 will suffer relatively more "damage" and phase perturbation than OAM+4 when each is impaired by the same obstruction at the same location in their common near field. That would correspond to the latter beam having greater self-healing ability than the former, lower-order OAM beam. If even partial self-healing does not occur, then our corresponding null hypothesis is that the same damage and phase-gradient perturbations will be consistently nearly equal on the two OAM beams.

To test our hypothesis, the OAM LOS link was set with the receiving array at 50m from the transmitting OAM antenna, such that the RX linear 4-element array was approximately tangent to the bottom edge of the (up-tilted) OAM conical beams. A data stream was modulated onto the RF carrier and transmitted on beam OAM+2 with a clear LOS to the far-field receiver, which captured the signals received on all 4 antennas. Next, the same process was repeated with the modulated carrier transmitted on beam OAM+4 and receive signals were captured. Then a partial obstruction (first author on a step-ladder; see Figure 5) was positioned at 10m from the OAM antenna, in its near field, but off-axis. (Note for scale, the parking-space marking lines are ~2 – 2.5 m apart in the picture in Figure 5.) The above processes were repeated with the obstruction in place and corresponding RX signals being captured with OAM+2 and OAM+4 for comparison.

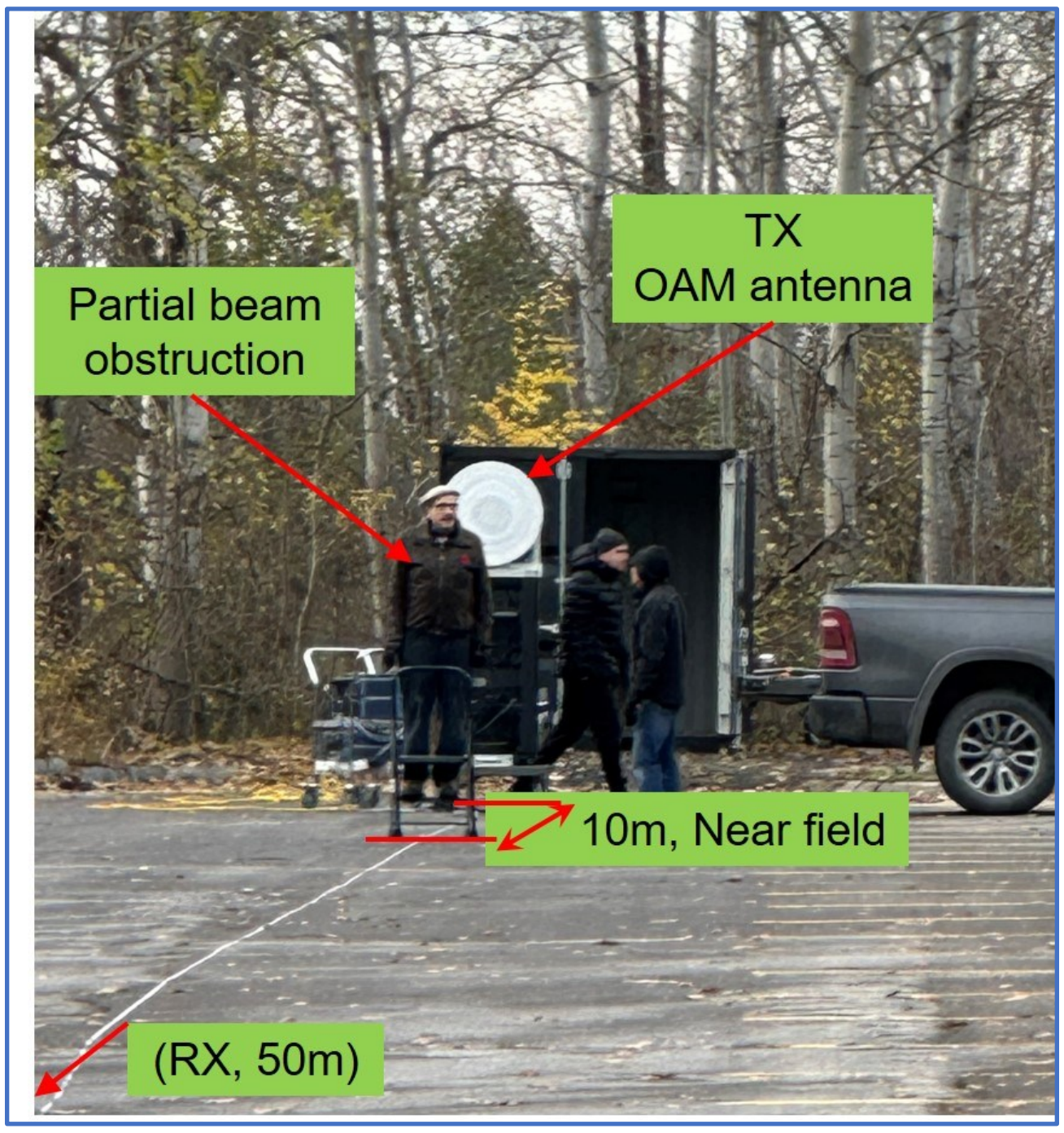

Figure 5. **Transmitting end of 50m long experimental LOS link with partial obstruction in near filed of transmitting OAM antenna.**

The receiving equipment was positioned in LOS in the far field at 50m from the OAM transmitter and aligned with the 1 degree cone angle of the OAM beams using a laser. It consisted of a linear array of four RHCP horn antennas, each connected to a 28 GHz receiver and coherent down-converter, which converted the signals coherently to baseband and captured them with synchronous digital-to-analog converters (DACs) on an evaluation system. The signals were post-processed in MATLAB, including maximal-ratio combining based on estimated channel-matrix pseudo-inverse. The receiving and signal-processing setup is shown in Figure 6.

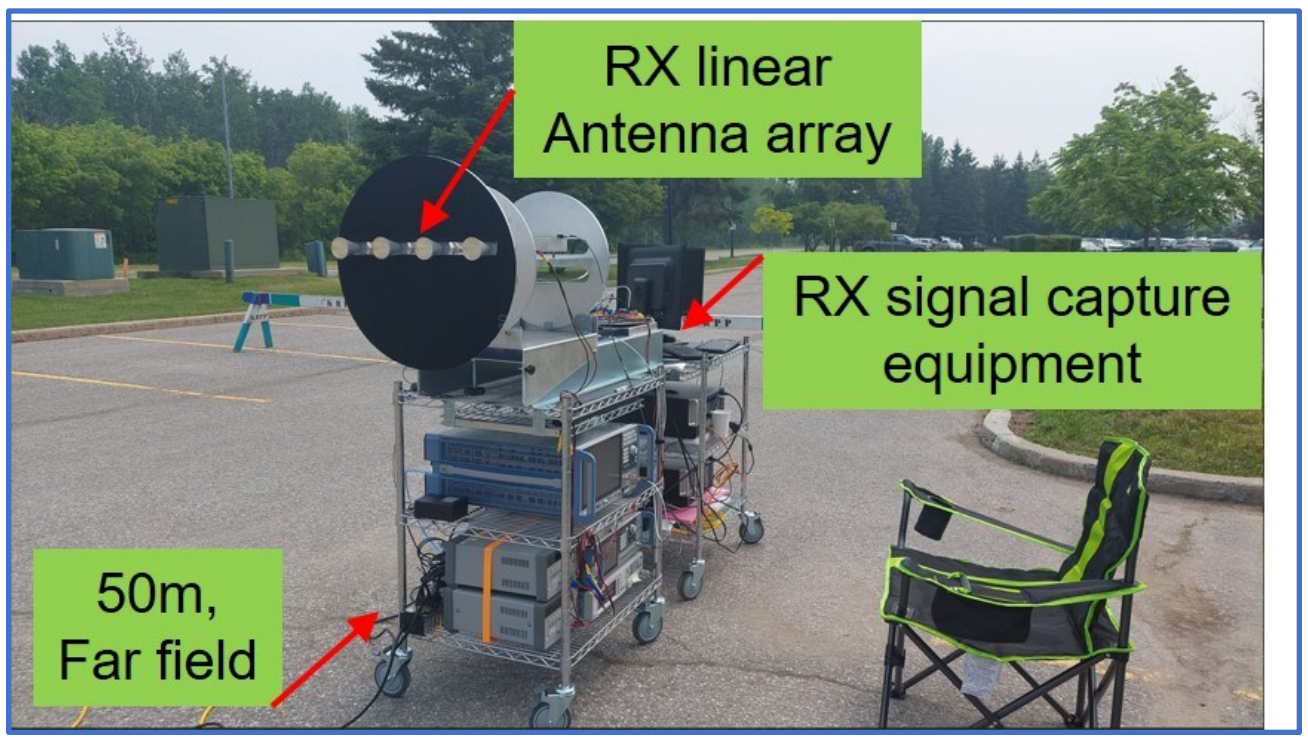

Figure 6. **Receiving end of experimental OAM outdoor link.**

The elements of the receiving antenna array in Figure 6 were spaced apart by 14 cm to span an aperture no larger than that of the transmitting antenna. They were connected to receivers denoted in linear order as RX1, RX2, RX3 and RX4.

Changes produced in the received signals by the partial obstruction were characterized for each OAM mode and are summarized in Table 2 below.

Table 2. **Average degradations of signal parameters by partial beam obstruction in OAM+2 and OAM+4**

| Parameter | RX Signal Power, dB | RX SNR, dB | RX Combined EVM, % | Propagation Channel RX #1 #2 #3 #4 Phases, ° |
|---|---|---|---|---|
| **Changes due to obstruction of beam OAM+2.** | | | | |
| **Change Averaged over 4 Receivers** | -6.05 | -4.5 | +2.74 | 0.0, 0.0, 0.0, 0.0 |
| **Changes due to obstruction of beam OAM+4.** | | | | |
| **Change Averaged over 4 Receivers** | -1.63 | -0.18 | +2.26 | +97.6, +98.6, 211.0, +93.1 |

It can be seen that the average signal parameters were less degraded by the same partial beam obstruction in OAM+4 relative to the greater degradation in OAM+2. Almost no impact on the pre-combining RX phases was observed on OAM+4. These findings are consistent with our hypothesis that at least partial self-healing is occurring on each OAM beam, and is more effective on the higher-order OAM+4 beam than on OAM+2 beam.

A more graphic indication of the superior self-healing in beam OAM+4 relative to that in beam OAM+2 is evident in the differences between the correlation functions of pilot signals in Figure 7 and Figure 8. The same pilot signals were transmitted on each OAM beam one at a time, and captured on each of the four RX antennas were correlated with the source pilot signal. The correlation magnitudes for the clear line-of-sight (LOS) and partially-obstructed LOS were plotted together in Figure 7 for beam OAM+2 and in Figure 8 for beam OAM+4.

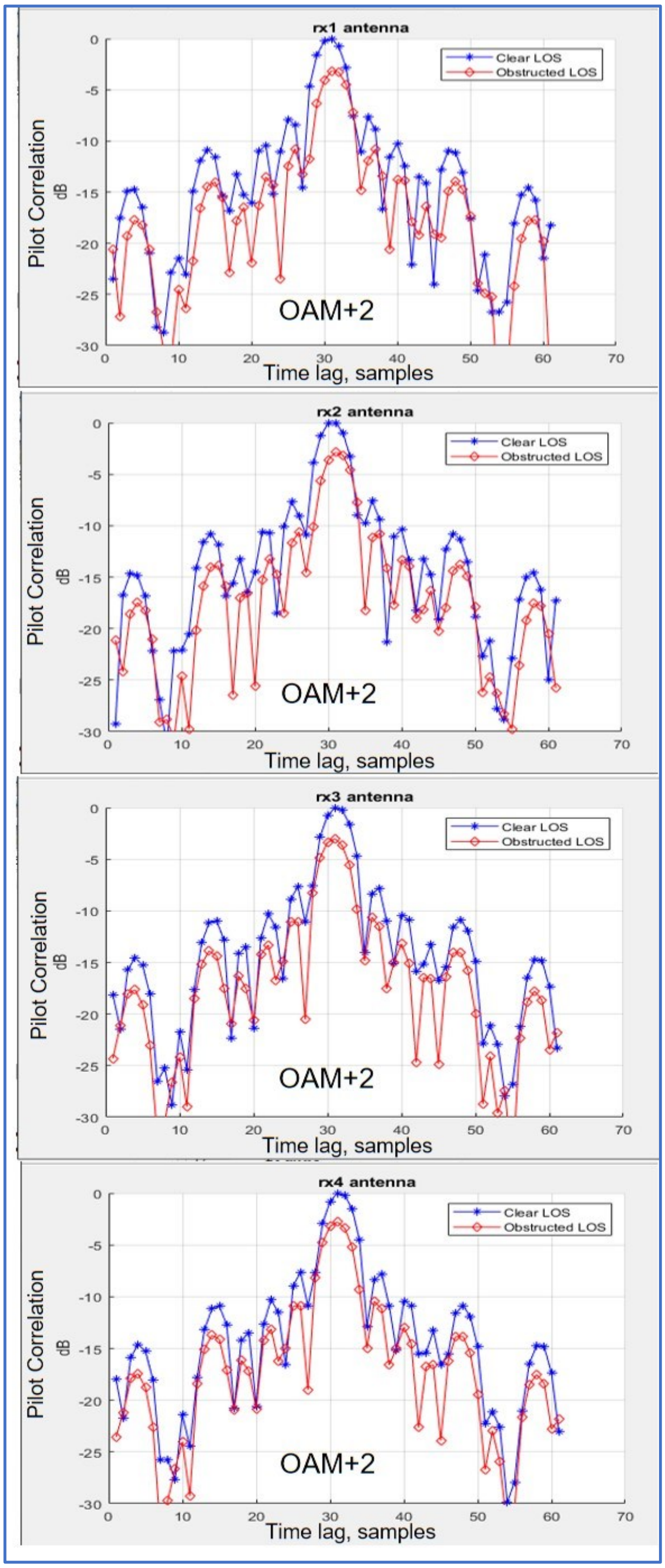

Figure 7. **Effects of partial obstruction of beam OAM+2 on the correlations of pilot signals received on each RX antenna.**

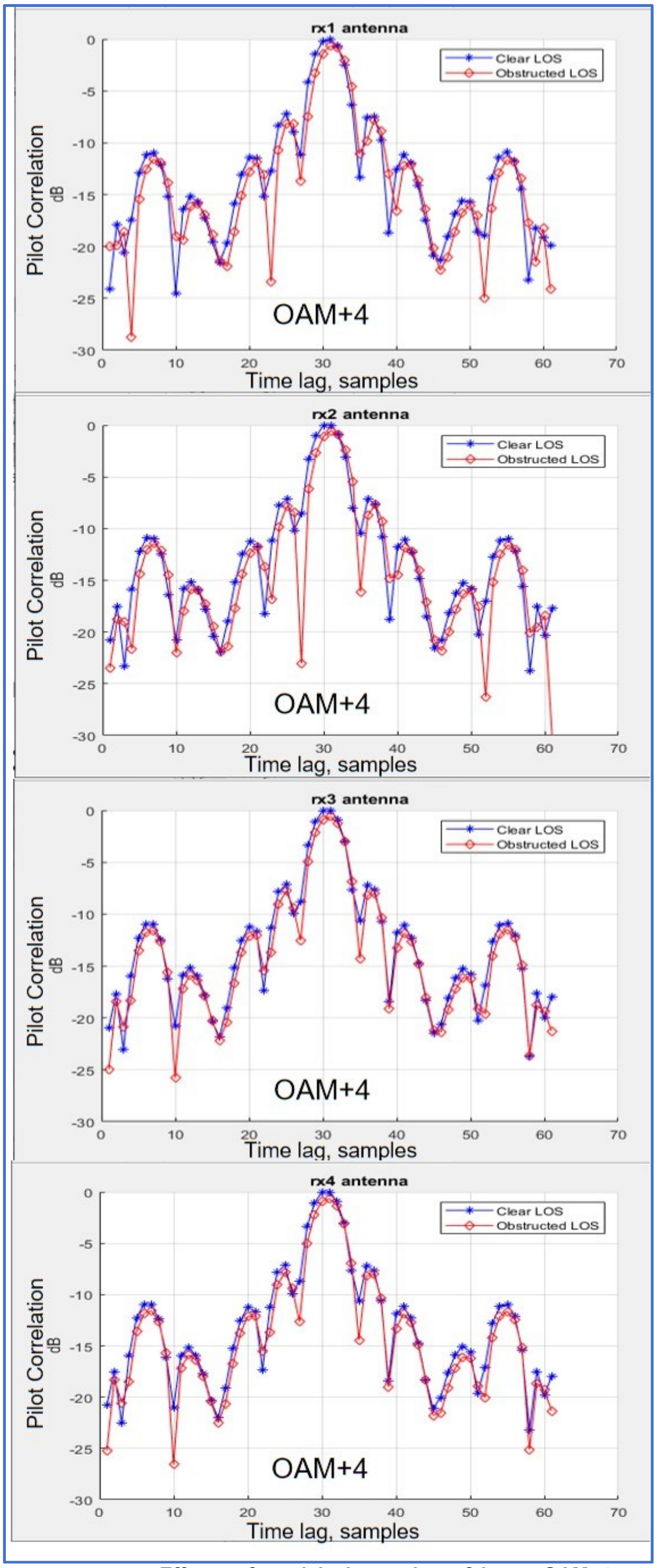

Figure 8. **Effects of partial obstruction of beam OAM+4 on the correlations of pilot signals received on each RX antenna.**

The pilot signal in each case was a pseudo-BPSK (pseudo-random symbols {(1+j), (-1-j)} at 30.72 Msymbols/s) signal sampled at 122.88 Msamples/s.

By inspection of Figures 7 and 8, it is clear that the degradations of the pilot correlations with the pilot signals received on beam OAM+2 caused by the partial obstruction, were greater than the corresponding degradations observed on beam OAM+4, caused by the same obstruction. Although the self-healing is never complete [5], some impact of the propagation channel does occur on both OAM beams. The observed fact that there is less impact on the higher-order OAM beam than on the lower-order one, indicates that self-healing is taking place in accordance with our hypothesis.

Because the beam-widths of both OAM beams were arranged to be equal and the signal strengths at the RX site were also adjusted to be equal, the differences in the propagation channels were dominated by the phase structures of the OAM beams. Consequently, these differences were the chief causes of the different degrees of self-healing effected by the phase structures of the OAM beams, when they experienced the same partial obstruction. This is evidenced by the greater residual effect observed in the pilot correlations on beam OAM+2 as compared to that observed on beam OAM+4.

Equivalently, migrating a given MIMO transmission in the presence of partial antenna obstruction from beam OAM+2 to beam OAM+4 provides some noticeable degree of self-healing from the effects of the obstruction, thus improving the quality of the spatial channel. This finding supports our initial hypothesis based on the relevant literature.

## IV. A PHYSICAL MODEL OF OAM SELF-HEALING

Recall from Section I that the equi-phase contours of an OAM beam form spirals around the beam $z$-axis. The Poynting vector is then locally orthogonal to these phase-fronts and also forms spirals around the beam $z$-axis, as shown in Figure 9. The free-space wavelength is associated with the wave vector $k = 2\pi/\lambda = 2\pi c/f$, which points in the same direction as the Poynting vector.

Note that the local free-space wave vector at some location, $\{z, R\}$ is not parallel to the beam axis, but is offset in the direction tangential to the cone of the OAM beam by a vector $k_T$. It will shortly be shown with the help of Figure 10 and reference [8], that this tangential wave-vector and the axial wave vector along the $z$ axis form a Pythagorean right triangle with the free-space wave-vector as

$$k^2 = k_z^2 + k_T^2 \quad (2)$$

In the treatment of reference [8], an additional wave vector in the radial direction, $k_r^2$, is added in the right side of (2), but for simplicity, here we will ignore this component by assuming the local shape of the OAM beam at $z = L$ is cylindrical rather than conical. A portion of such local cylinder, corresponding to one $2\pi$ cycle of electrical phase, is shown unrolled at the bottom of Figure 9.

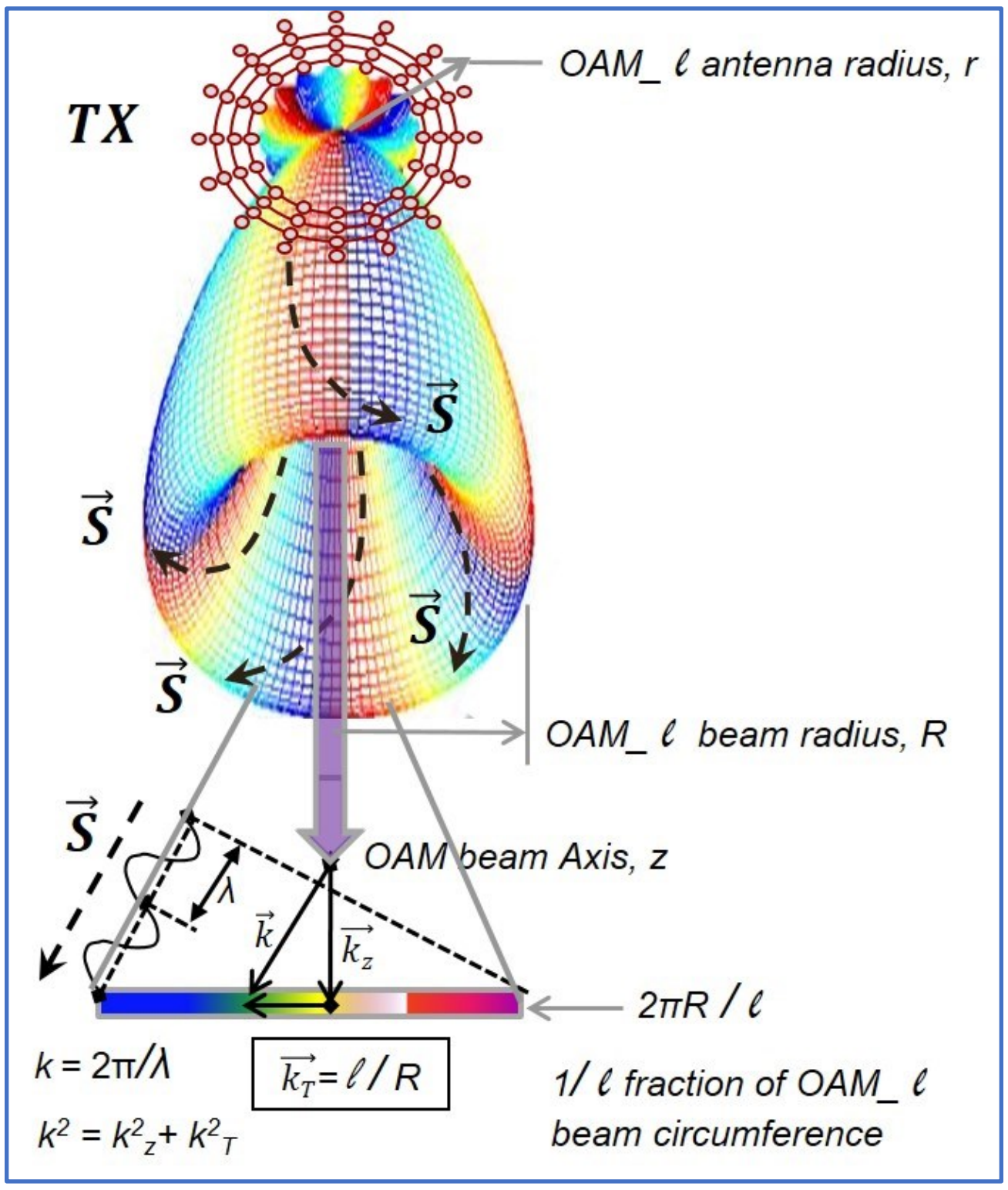

Figure 9. **Poynting vectors and wave vectors associated with OAM beam, with simplified details in vicinity of a fixed location on beam's z-axis, where its cone has radius *R*.**

It will be appreciated that the radius $R$ increases with $z$ as one proceeds along the beam axis into the far field, where at $z=L$, we have approximately $sin(\theta) \sim tan(\theta) \sim R/L$ in equation (1). Taking the peak of beam OAM_$\ell$ to be at approximately $x=\ell+1$ in Figure 4, leads to

$$R(z = L) \approx \frac{L(l+1)}{rk} \tag{3}$$

where $r$ is the radius of the transmitting OAM circular antenna array, $k$ is the free-space wave vector and $\ell$ is the order of the OAM mode (scaled to OAM+2 per Figure 4).

With the aid of Figure 9, one way to think about the self-healing property of OAM beams is that the Poynting vector "winds around" the partial obstruction as the OAM wave-front propagates, due to the tangential component of the wave vector, $k_T$. Note that the wave vectors relate to phase velocities in their respective directions. As the beam propagates, this tangential wave vector $k_T$ moves the phase-fronts around the periphery of the conical main lobe of the OAM beam. The partial obstruction interrupts this tangential progression by blocking some of the phase-fronts, but the successive ones that are not blocked, act as Huygens sources and propagate into the shadow of the partial blockage. The more phase-fronts per circumference of the OAM beam, the more successful they will be in filling the shadow region. That is intuitively one reason why higher-order OAM beams tend to self-heal better than lower-order ones.

We show next that this tangential vector responsible for the self-healing effect is directly proportional to the OAM order, and its magnitude is given by

$$k_T = \ell / R \tag{4}$$

Equation (4) then predicts that the self-healing effect is greater at smaller beam radii closer to the OAM antenna, and also increases with OAM order, $\ell$, which agrees with our hypothesis. (The afore-mentioned radial component of the wave vector also contributes to filling in the shadow region in general structured beams, but was ignored for simplicity in this exposition, because the OAM beams were not deliberately structured in the radial direction, being radiated essentially from a narrow circular ring of antenna elements.)

To complete this rationale, it remains to derive the wave vectors related in equations (2) and (4) above. To that end we employ reference [8] as it relates to equations (4) - (7) therein, beginning with the cut-off OAM mode number in a coaxial waveguide carrying a radially-polarized wave. This cut-off condition can be derived with the aid of Figure 10 beginning with the familiar relation for the cut-off wavelength of the lowest-order TE mode in a rectangular waveguide. To illustrate, Figure 10 depicts the evolution of the rectangular waveguide into a coaxial waveguide, with the perfect electrical conductor (PEC) boundary condition replaced by a phase-continuity condition, and the half wavelength replaced by a full wavelength (to preserve phase-continuity modulo 2π radians). The coaxial waveguide behaviour is supported by the analysis in reference [8].

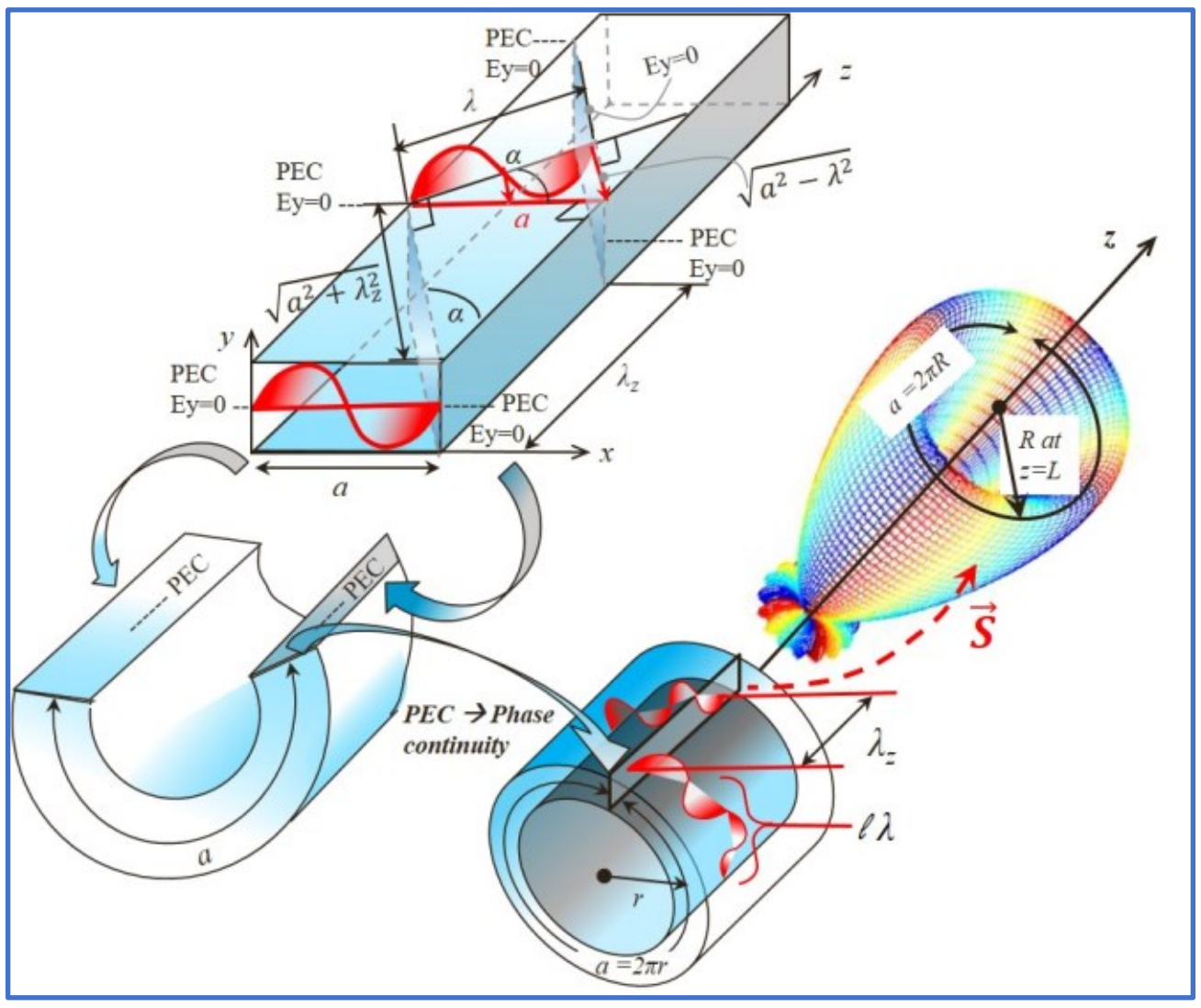

Figure 10. **Illustration of the derivation of the wave vectors comprising a local region of an OAM beam.**

We further evolve this behaviour to radiated OAM beams, proceeding from the cylindrical structure and radial polarization in the coaxial waveguide, to the conical beam structure and arbitrary polarization of the radiating OAM beam in Figure 10.

Recall that in the rectangular waveguide, the transverse electric fields ($E_y$) arrange themselves so that their zero-value nodes fall on the PEC boundaries along the *x*-axis (*x=0* and *x=a*) in the transverse-electric (TE) modes. This corresponds to the projection of the nodes from a plane-wave component having free-space wavelength $\lambda$ propagating in the waveguide at angle $\alpha$ to the x-axis, as shown for one wavelength in Figure 10.

$$\cos\alpha = \frac{\lambda}{a} \tag{5}$$

$$\sin\alpha = \frac{\sqrt{a^2-\lambda^2}}{a} \tag{6}$$

$$\tan\alpha = \frac{a}{\lambda_z} = \frac{\sin\alpha}{\cos\alpha} \tag{7}$$

It also corresponds to the guided-wavelength along the *z*-axis, $\lambda_z$, which can be determined from the geometry using (5)-(7) above, as

$$\lambda_z = \frac{\lambda}{\sqrt{1-\left(\lambda/a\right)^2}} \tag{8}$$

We have used the projection of the zero-$E_y$-field at one free-space plane-wave length instead of the one at half of that wavelength because we want to preserve phase continuity in the subsequent development.

Proceeding down from the rectangular waveguide in Figure 10, we visualize bending the broad walls around the z-axis so that the short walls with the PEC boundary condition come toward each other along the circular path of the bend, as in the lower left of Figure 10. As we continue bending the waveguide, the "*a*" dimension forms a circle of circumference *r=2π/a* and the PEC boundary walls merge together. Then the top broad wall becomes the inner cylindrical conductor and the bottom broad wall becomes the outer conductor of a coaxial waveguide, and the merged short walls merge the zero-field nodes of the transverse electric field with phase continuity. We can now substitute $a=2\pi r$ for the broad wall width in and any number of whole free-space wavelengths, $\ell\lambda$, for $\lambda$ in equation (8) to obtain the guide wavelength for a coaxial waveguide of radius *r* as

$$\lambda_z = \frac{\lambda}{\sqrt{1-\left(\frac{\ell\lambda}{2\pi r}\right)^2}} \tag{9}$$

which agrees with equation (6) of reference [8] for the OAM cut-off wavelength. To see this in terms of wave vectors, we next square both sides of (9) and isolate the free-space wavelength in terms of the guide wavelength. At the same time we migrate from the coaxial waveguide to free space at *z=L*, where we have the OAM beam radius *R* in place of waveguide radius, *r*. Since *R* increases with *z*, there is no longer an OAM more cut-off in free space. This results in

$$\lambda = \frac{\lambda_z}{\sqrt{1+\left(\frac{\ell\lambda_z}{2\pi R}\right)^2}} \tag{10}$$

which we then proceed to express in terms of wave vectors defined as $k=2\pi/\lambda$ and $k_z=2\pi/\lambda_z$.

$$k = \frac{2\pi}{\lambda} = \frac{2\pi}{\lambda_z}\sqrt{1+\left(\frac{\ell\lambda_z}{2\pi R}\right)^2} \tag{11}$$

Squaring both sides of (11) allows to rewrite it as

$$\left(\frac{2\pi}{\lambda}\right)^2 = \left(\frac{2\pi}{\lambda_z}\right)^2 + \left(\frac{\ell}{R}\right)^2 \tag{12}$$

which is recognized as equation (2) with the correspondence of the tangential wave vector as defined in equation (4). It is this tangential wave vector in equation (4) which is responsible for most of the self-healing effect observed in the relative results of the experiments described in this paper.

## V. CONCLUSIONS AND FUTURE WORK

This paper describes for the first time the effects of self-healing of OAM beams in a mm-wave experimental communications link. The effects were observed in relative terms on two different OAM beams which were arranged to be equivalent in their beam parameters except for their helical phase structures, i.e. their orders, $\ell$, of OAM which was expected to be responsible for the different amounts of self-healing observed. The beams were received one at a time on a 4-antenna MIMO experimental receiver and captured for post-processing. Although the beam structures were not amenable to direct measurement as in experiments performed by prior researchers, we measured communications link and channel parameters such as SNR, EVM, channel phases from OAM transmitter to each MIMO receiver and correlations of received pilot signals with the source pilot signal. Impact of a near-field partial obstruction of the transmitting OAM antenna were noted for OAM+2 and compared with the impacts of the same obstruction on OAM+4. The impact of partial obstruction was greater on OAM+2 than on OAM+4, indicating that more self-healing occurs on the latter, higher-order OAM beam, which agrees with our hypothesis. A physical model of the self-healing mechanism at work here was described, which led to the hypothesis that if self-healing was observed, it would be greater on higher-order OAM beams, and manifest as a greater impact of the partial obstruction on the lower-order OAM beam.

The physical model does not predict any quantitative effects of self-healing or obstruction, but it does agree with analyses and experiments described in prior literature, mostly in the optical field. In future work we plan to apply various MIMO combining algorithms and further characterize the robustness of multiple simultaneous OAM beams to partial obstructions.